\documentclass{elsart}

 \usepackage{epsfig}

\usepackage{amssymb}
\begin{document}
\begin{frontmatter}


%\title{Versione finele del 4 marzo}

\title{Metastable states, anomalous distributions and correlations in the HMF model}

\author%[label1]
{Alessandro Pluchino},  \ead{alessandro.pluchino@ct.infn.it}
\author%[label1]
{Vito Latora}, \ead{vito.latora@ct.infn.it}
\author%[label1]
{Andrea Rapisarda} \ead{andrea.rapisarda@ct.infn.it}

\address%[label1]
{Dipartimento di Fisica e Astronomia,
Universit\'a di Catania,\\
and INFN sezione di Catania, Via S. Sofia 64,  I-95123 Catania, Italy}

\begin{abstract}
We study  
the microscopic dynamics of the metastable Quasi-Stationary States (QSS) in
the Hamiltonian Mean Field (HMF) model, a Hamiltonian system of N
classical inertial spins with infinite-range
 interactions which shows a second order phase 
transition. In order to understand the origin of
 metastability, which appears in an energy 
 region below the critical point, 
we  consider two different classes of out-of-equilibrium initial
conditions, both  leading to QSS, and  having
 respectively initial  magnetization equal to one (M1 IC) 
and equal to zero (M0 IC). We   
compare the corresponding $\mu$-space, the  resulting
velocity pdfs and correlations, and the eventual aging features of
the microscopic dynamics.
In both cases the model exhibits non-gaussian pdfs, though
anomalous correlations are present only when the system
is started with  an initial  magnetization equal to one.    
In the M0 IC case the relaxation to equilibrium is almost
exponential, while, for M1 IC,  
when correlations and aging are found, the decay is a power-law
and the overall behavior can be very well reproduced 
 by a Tsallis q-exponential function.
These results  contribute to clarify 
the overall scenario, which is more complex than previously
expected and stress the importance of the dynamics in the relaxation
process.
The  nonextensive statistical mechanics formalism proposed by Tsallis seems 
to be valid, in the out-of-equilibrium phase, 
when correlations and strong long-term memory effects 
emerge. This regime becomes stable if the $N\rightarrow \infty $ limit is performed
before the   $t\rightarrow \infty $ limit.
\end{abstract}

%Paper submitted to Physica D - [cond-mat/0303081]
Revised version 01/04/2003 - Physica D in press

\end{frontmatter}

\section{Introduction}
\label{int}
There is by now a huge number of papers devoted to 
understanding the physical foundation and the possible 
applications of the nonextensive statistical mechanics formalism
proposed by Constantino Tsallis in 1988 \cite{tsa1}. 
In particular, although the topic is still a matter of debate \cite{cho}, 
the last few years have seen an important turning point for what concerns
dissipative
systems both low-dimensional (for example unimodal
maps) \cite{lyra,logvito,fulvio}
and high-dimensional ones 
(for example turbulence, plasma and nuclear physics, econophysics) 
\cite{beck,cohen,bc,bog,wilk,borland}.
Progress has been made also for conservative systems 
both for maps \cite{stand} and for high-dimensional 
hamiltonian systems with long-range interactions 
\cite{lat0,lh02,celia,gian1,cab,monte,nobre}, but 
open problems still remain and need a more detailed investigation. 
With this scenario in mind, in this paper we present and discuss 
new numerical simulations relative to the Hamiltonian Mean Field
(HMF) model, 
a system of $N$ fully coupled classical spins which has been intensively 
studied in the last years.
It is nowadays well known\cite{lat0,lh02,ant1,lat1,lat2,lat3,ant2,ant3,lat4} 
that, for some values
of the initial energy, below the critical point, the system does not
immediately relax to the Boltzmann-Gibbs equilibrium, but remains trapped
in anomalous metastable states and exhibits non-Gaussian velocity
distributions
that can be reproduced  by the probability density functions (pdfs)
predicted by the Tsallis nonextensive thermodynamics \cite{tsa1}.
Such states have been named Quasi-Stationary States (QSS) because they
become 
stable in the thermodynamic limit if the $N\rightarrow \infty$ limit is 
taken before the $t\rightarrow \infty$ limit. 
In fact in such  a case the force 
between spins tends to zero  
and  the largest Lyapunov exponent vanish \cite{lat0,lh02}.

In this work we study two classes of out-of-equilibrium 
initial conditions that lead to QSS, and we discuss how the different  
initial conditions may affect the dynamics of the relaxation 
towards equilibrium. 
In particular, we concentrate our attention on the existence of
correlations 
in the orientation angles and velocity of the spins, on the presence of 
anomalous velocity distributions, and on the 
clusters formations. 
The main result of this paper is that Tsallis nonextensive 
thermodynamics formalism seems to provide a general framework, 
to interpret the statistical properties of the dynamically 
created QSS, for  Hamiltonian systems with long-range interactions
\cite{tsa1}, 
only when long-range correlations and fractal structures in phase space 
are present. 
Recently some critical comments \cite{zan} have been raised on a previous
study 
of two of us \cite{lat0}. In this paper we provide 
new calculations and detailed discussions which 
reply in part to this criticism and clarify also some points 
advanced in another work \cite{yama}. A more detailed reply  to the 
points raised in ref. \cite{zan} will be presented elsewhere \cite{reply}.

The paper is organized as follows. In section II the details of the model
are summarized and the initial conditions used are discussed. 
Numerical results for QSS are discussed in section III, while
velocity pdfs and  dynamical correlations are studied in a quantitative
way
in section IV. Conclusions are drawn in section V.

\section{Model and initial conditions}
\label{model} 
The HMF model, consists  of
N planar classical inertial spins interacting through an
infinite-range potential \cite{ant1}. The  Hamiltonian can be written as:

\begin{equation}
\label{h}
        H= K+V
= \sum_{i=1}^N  {{p_i}^2 \over 2} +
  {1\over{2N}} \sum_{i,j=1}^N  [1-cos(\theta_i -\theta_j)]~~.
\end{equation}

In the latter $\theta_i$ is the orientation angle of the $ith$ spin 
 and $p_i$ the
conjugate variable representing the rotational velocity, considering the
mass of the rotators 
equal to one. The summation in V is extended to all couples of spins
and is not restricted to first neighbors.
In order to make H formally extensive, i.e.
$V\propto N$ when $N\rightarrow\infty$\cite{tsa1,celia},
the coupling constant in the potential is
divided by N. However, this often adopted prescription cures only a part
of the problem since
the energy remains non-additive, that is    the system cannot
be simply divided in two independent sub-systems.
The magnetization
M, i.e.  the modulus of  ${\bf M}={\frac{1}{N}} \sum_{i=1}^N {\bf m}_i$,
whith  ${\bf m}_i=[cos(\theta_i), sin(\theta_i)]$, is the order parameter
of the 
model. The latter   shows a
a second-order phase transition from a low-energy
clustered  phase with magnetization  $M\sim 1 $,
to a high-energy one, where the spins are homogeneously oriented
on the unit circle and $M \sim 0$.
 The model can be solved exactly in the canonical ensemble and the
dependence of the energy density $U = E/N$ on the temperature $T$,
is  given by \cite{ant1,lat1}

\begin{equation}
U = {T \over 2} + {1\over 2} \left( 1 - M^2 \right) ~.
\end{equation}

The critical point is at energy density $U_c={3 \over 4}$
corresponding to a critical temperature $T_c={1 \over 2}$ \cite{ant1}.
The dynamics of HMF model can be investigated by starting the
system with out-of-equilibrium initial conditions
and integrating numerically the equations of motion \cite{lat1}.
In a special region of energy values (${1\over 2}<U<U_c$)
the results of the simulations show, for a transient regime which depends
on the system size, a disagreement with the canonical ensemble.
In this region the dynamics is characterized by  L\'evy walks and
anomalous diffusion, while  in correspondence
the system shows a  negative specific heat \cite{lat3}.
Ensemble inequivalence and negative specific heat
have also been found in self-gravitating systems \cite{thir,lyn},
nuclei and atomic clusters  \cite{gro,dago,ato},
though in the present model such anomalies emerge as dynamical metastable
features  \cite{ant2,lat4} and vanish, if $N$ is finite, for 
$t\rightarrow \infty$. 
In this paper we focus on a particular energy value belonging to
the anomalous region, namely $ U=0.69$, and we study the
time evolution of temperature, $\mu$-space structures, 
dynamical clusters in angle,
velocity correlations and velocity
distributions considering two out-of-equilibrium initial conditions
in particular:
\begin{itemize}
\item  the first one, used  already in previous papers,
consists of  all orientation angles $\theta_i=0$
and velocity $p_i$ uniformly distributed accordingly to the total energy
available. In this case the potential energy is initially zero and
we have the maximum of kinetic energy, while
the total magnetization M is equal to 1;
\item in the second case considered, both   
 angles and velocities are  uniformly
distributed. This implies that the total
magnetization 
M is initially
zero, so that  the potential energy is  $1\over2$, see eq.(2). Therefore
in this case 
 only a smaller fraction of total energy is left for the  kinetic part.

\end{itemize}
For both initial conditions the total momentum is initially put to zero.
In the following, for brevity, we will indicate these two initial
conditions respectively by $ M1$ and $ M0$, thus referring to the initial
value of
the magnetization.

%%%%%%%%%%%%%%%%%%%%%%%%%%%%%%%%%%%%%%%%%%%%%%%%%%%%%%%%%%%%%%%%%%%%%%%%%

\section{Metastable equilibrium and Quasi-Stationary States}

We discuss in this section the numerical results obtained starting
from M1 and M0 initial conditions. Different sizes of the system are
studied. 
The accuracy of the calculations
is such that one gets an energy conservation $\Delta E / E = 10^{-5}$.
As in previous papers the fourth order
Yoshida simplectic algorithm \cite{yoshi} has been used for the
integration.
At variance with
the previous publications the averages where taken only over the ensemble
and not also 
along the  time evolution. This point will be discussed in detail later.

In fig.1 we plot, for system  sizes N=500 and N=1000, the time evolution of
twice the kinetic energy
per particle, a quantity
which is the  temperature in stationarity situations. 
Different regimes exist according
to the initial conditions  considered. More specifically, when  starting
with $ M1$
initial condition (IC) four regimes are
observable:
\begin{enumerate}

\item  a fast decay from $M=1$ to $M\sim0$ in a time which is about
80-100. No sensible
       dependence on the size is observed;

\item  a plateau region with $M(N)<M_{eq}$ and  $T(N)<T_{eq}$
        whose duration depends also on the size of the system as N
\cite{lat0};

\item  a power-law relaxation towards the true equilibrium value
$T_{eq}=0.476$  \cite{lh02};

\item  an equilibrium regime with $M_{eq}=0.307$.

\end{enumerate}

In refs.\cite{lat0,lh02} the kinetic energy per particle was actually computed
considering
not only an ensemble average, but also a time average. This was done to 
improve statistics,  though   in principle time averages are justified only 
when correlations do not exist.  
 In fig.2 the old procedure  and the new one are compared.  
One observes that  the two
methods give the same  plateau if a  large
number of
events are considered in the ensemble  average.
The only difference is in the  decay from the initial condition to the QSS
regime. One can easily check that also the scaling laws found in
ref. \cite{lat0}
continue to be valid. In particular the lifetime of the plateau  diverges
with $N$
and the value of the plateau temperature $T_{QSS}(N)$ 
converges towards the asymptotic temperature
$T_{\infty}=0.38 $ as $N\rightarrow \infty$ following the same laws 
of refs.\cite{lat0}.
At variance with the criticism advanced in \cite{zan},
it is evident that one can really  talk of an approximate plateau for the 
average 
kinetic energy per particle, see fig.2. The log timescale used in fig.1 
and in previous papers \cite{lat0} is
useful to have an overall view of the temporal behavior 
and does not change the macroscopic quasistationary
features.
This will be clearer when analizing also the M0 IC, for which the system
is initially
put in the state at $M\sim 0$ and $T\sim 0.38$ and there it remains for quite a
long time which also in this case depends on the size studied.
%%%%%%%%%%%%%%%%%%%%@@@@@@@@@@@@@@@@@@@@@@
In fact when starting with $M0$ IC only three regimes are
observable:
\begin{enumerate}
\item   a plateau region with $M\sim 0$ and $T\sim 0.38$ whose length
          depends on the size as $ N$. This has been checked numerically
	  but it is not shown here for lack of space;
\item   a power-law relaxation towards the true equilibrium value, but
         faster than the previous $M1$ case;
\item   an equilibrium regime with $M_{eq}=0.307$.
\end{enumerate}
In general, for both IC,  the values obtained for 
the magnetization and the temperature in 
the plateau regime for $N\rightarrow \infty$
 are consistent with the 
{\it caloric curve}
 (2) extended below the critical point. Therefore this average 
procedure is a-posteriori justified  and consistent with a
metastable  macroscopic thermodynamic
description, contrary to what claimed in ref.\cite{zan}. 
Although both initial conditions drive the system
 towards the same out-of-equilibrium 
macroscopic values
for the magnetization and the temperature in the $N\rightarrow \infty$ limit,
for  the QSS regime, the corresponding 
 microscopic  dynamics is very much different in the two cases.  
This fact can be
seen when one plots the time evolution of the $\mu$-space, as done in
fig.3 for
a typical event and  $N=10000$.
In red we represent the  
 rotators having at time $t=0$ the highest velocities.
In this way it is possible to visualize the kind of
mixing occurring for  the initial conditions considered.
It is immediately evident that there are
correlations and structures in the plateau regime only for M1 IC, while
for M0 IC the distributions remain homogeneous with a very slow mixing of
the
initially fast rotators.
This different behavior can be observed also in figure 4,
where we plot the pdf in angle at various times. 
A typical realization for N=10000
is plotted. 
One observes clusters formation in the plateau regime only for
M1 while these are completely absent for M0 IC. However, at equilibrium,
the pdfs are
equivalent and the plots, for both M1 and M0, show only one rotating big
cluster
corresponding to $M=0.307$ which is the equilibrium magnetization
(according to  the canonical  prediction eq.(2) with $U=0.69$ and $T=0.476$).

These numerical results reinforce our previous statement that 
although macroscopically similar the two metastable plateaux 
have  a very different microscopic dynamics. Further evidence is 
provided if one studies the entropy and free energy as a function of time.

 This is done in figs.5 and 6 respectively.
To obtain the entropy we adopted a lattice over the $\mu$-space made up of
$n=10000$ cells and we used the following formula:

\begin{equation}
\label{h}
        S(t)
= < -\sum_{k=1}^n  {f_k(t) ln(f_k(t))} >,
\end{equation}

where $f_k$ is the relative frequency of points for the $kth$ cell. The
brackets $<...>$
denote ensemble averages. The curves plotted in the figures refer to an
ensemble of
$1000$ events for $N=1000$. 
This  particular size was used in this case and in
other calculations
when a good average over the ensemble was essential  in order to have a
robust
value, computer time being   too long for  greater
sizes and such big ensemble averages. 
On the other hand 
no qualitative changes are anyhow  observed for greater sizes.
As intuitively expected from the temporal evolution of the $\mu$-space,
 the entropy grows with time towards equilibrium only for M1 IC
due to the high degree of order of this initial configuration,
while it remains fixed at the maximum value for the 
other, much more uniform, initial condition,
$M0$. Correspondingly, we compute the free energy  by extrapolating 
to a dynamical situation the
equilibrium thermodynamic formula

\begin{equation}
F(t) = U - T(t) S(t) ~~.
\end{equation}

As shown in fig.6 the free energy remains constant for $M0$ in the QSS
regime.
This implies that for this IC the system is consistently in a metastable
regime,
which is not however the true one that is  reached only at greater times,
when the free energy of the system has a  real global minimum.
 On the other hand,
for the $M1$ IC, the
figure shows a  different temporal evolution before equilibration. 
After a 
short rapid transient decay,   an almost  flat
region, at a different free energy value,   always emerges in the QSS
regime, 
implying also in this case the presence of 
a metastable situation.
 The use of formula (4) is thus a-posteriori justified since the
temperature is almost constant in the plateau regime. It is also very 
 useful in clarifying  the scenario
that emerges which is consistent with a quasi-equilibrium situation. 
This results reinforce the 
validity of our
procedure to extract average thermodynamic quantities from microscopic
dynamics and can
be considered a  further answer to the criticism raised in ref.\cite{zan}.
On the other hand, 
 the fact that there is no plateau for the temporal evolution of
the entropy for $M1$ IC
reflects  the slow microscopic dynamics  observed in the evolution of the
$\mu$-space structures.
This is perfectly  in line with the quantitative analysis performed by
means of the
correlation dimension in ref.\cite{lat0}. Thus one can summarize this
situation
by saying that only at a macrosopic level the quasi-stationarity is valid,
while microscopically
the system never stops its slow temporal evolution. The evolution really stops
only for $N=\infty$, since in this
 case the magnetization in the QSS is exactly 
zero and so is the force which is proportional to $M$.
In this sense the inversion of the time limit with the size limit to infinity
stabilize the QSS regime. 
However,  when considering finite 
$N$ systems, the fluctuations due to
finite-size noise, of
the order $\sim 1/ \sqrt N$, induce  a force which is not exactly zero, and when they 
  grow with time,  the system is slowly forced 
towards equilibration. 
As shown in ref.\cite{lat0} the accuracy of our calculations is much smaller than this
finite-size noise.

For the M0 IC  no entropy increase is observed. But also in this case
some microscopic
dynamics exists. This is proved by the mixing of fast particles in
fig.3. However no holes or
structures are 
present in this case and this is the reason why  the entropy temporal
evolution is almost flat.

It is important to stress the fact that, for very long times, both
the two initial conditions  lead the system towards  the equilibrium
expected values.

We would like to notice at this point that the slope of the 
entropy growth in fig.5
does not give the Kolmogorov-Sinai entropy as demonstrated numerically for 
maps \cite{logvito,bara}. In fact  in our case we are considering the reduced 
$\mu$-space and not the complete phase-space  which has to be used. 
Unfortunately in our case the phase space has too many dimensions 
and this fact prevents the direct investigation of the generalized 
q-entropy proposed by Tsallis along the same lines of ref.\cite{logvito}
for maps.

\section{ Anomalous velocity pdfs, velocity correlations and
aging  }

In this section we study in a quantitative way the eventual presence of
 velocity correlations
and memory effects  in the HMF dynamics.
In Fig.7 we report the velocity pdfs for the plateau regime (at $t=1000$)
and at equilibrium  $(t=1000000)$  and
for the case
 N=10000.
For both IC,  the initial velocity pdfs quickly acquire and maintain
during the entire duration of the metastable state a  {\em non-Gaussian
shape}.
But while for M1 IC they can be fitted with a truncated generalized
Tsallis pdf
\cite{lat0},
this is not possible for M0 IC. In fact in this case
the pdf shows a flat almost rectangular shape,
 which clearly reveals the fact that structures
and correlations, present in the other case, are here missing.
Finally, for very long times the system reaches always the equilibrium
distribution and the agreement with the gaussian is almost perfect.

A quantitative way to estimate the velocity correlations is the
calculation
of the two-time autocorrelation function, defined as
\begin{equation}
{ C}(t,0)= {< {{{\bf P}(t) \cdot {\bf P}(0)}}> -
{{<{\bf P}(t)>\cdot <{\bf P}(0)>}} \over {\sigma_p(t)\sigma_p (0)}}~~,
\end{equation}
where ${\bf P}=(p_1,p_2,...p_N)$ is  the   N-component  velocity vector
and the brackets $<...>$ indicate the
average  over the ensemble, while  $\sigma_p(t)$ and $\sigma_p(0)$
are the standard deviations at time t and at time zero.
In fig.8 we plot the velocity autocorrelation functions
for M1 and M0 IC in the QSS  regime, obtained for the case N=1000 and
an average over 500 events.
It is immediately evident that for M0 the autocorrelation decays faster
than for M1 IC. This result clarifies the controversial   numerical
results
 obtained  in ref.\cite{yama}  by  using a
slightly different, but equivalent, definition for the autocorrelation
function.
In fact we do  find an almost exponential behavior only for the M0 IC, while in
ref. \cite{yama}
the author claimed to find an exponential in both cases. In general
we  can reproduce the autocorrelation decay by means of the  q-exponential
function
\begin{equation}
e_q(x)= A {\left[  1-(1-q) {x \over \tau} \right]} ^{1\over(1-q)}
\end{equation}
proposed by Tsallis in his thermodynamic nonextensive q-formalism
\cite{tsa1}, where  $A  $ is the saturation value
 and  $\tau$ the characteristic time. One gets the usual exponential for $q=1$.
In this way
 we can quantitatively discriminate  between the two different initial
conditions.
In fact we get for M1 the value $q=1.55$, revealing a power law scaling,
while for M0 we get a value of q much smaller, i.e. $q=1.12$. This
 means that for M0 the decay tends to be exponential (an exact exponential
is
obtained when $q=1$). This almost exponential decay is also evident in the
inset
of the figure, where we show the same data
in linear-log scale. An exponential fit is also reported for comparison.
Such numerical results may be considered a quantitative confirmation
of the different microscopic dynamics of the system according to the
particular
initial condition,  in spite of the
macroscopic presence of QSS for both  cases.

%%%%%%%%
%%%%
%    aging
%%%%%%%%%%%%

It is interesting to investigate also if the  relaxation process observed 
in fig.8 
depends on the history of the system, i.e the possible existence of 
strong long-term memory effects,   
 the so-called {\it aging phenomenon}. This is typical of
glassy systems but it has also been found in a great variety of
complex systems \cite{stil,bou,cugli}. 
Recently it  has been  observed  also in
the HMF model  by Montemurro, Tamarit and Anteneodo  \cite{monte}.
In the following 
we perform a similar  analysis to study this effect for both M1 and M0 IC,
 but considering only  the  N-component
velocity vector defined previously. In their case also angles
were taken into account in the state vector entering the correlation
function and  only M1 IC were considered.
The autocorrelation function to be considered is that one defined  before
but calculated with respect to a set of four different waiting times $t_w$
(that plays the role of  {\it age of the system}) 

\begin{equation}
{ C}(t+t_w,t_w)= {<{{{\bf P}(t+t_w)\cdot {\bf P}(t_w)}}> -
{{<{\bf P}(t+t_w)> \cdot <{\bf P}(t_w)>}} \over {\sigma_p(t+t_w) \sigma_p(t_w)}}.
\end{equation}

In Fig.9 (a) we show the autocorrelation decay
for M1 IC, while in Fig.10 (a)that one for M0 IC.
If aging were not present, all the curves should overlap into one, because
the function $C(t+t_w,t_w)$ would depend only on the difference between
the two times
$t+t_w$ and $t_w$, i.e.  $C(t+t_w,t_w)=C(t)$,
and the decay would be exponential. Actually  this is not
what happens as shown in the figures.

For M1 IC aging properties surely  exist,
i.e. it can be observed a characteristic two times
dependence on $t$ and $t_w$ and  the data can be scaled onto one
curve, panel(b),
considering as discussed in ref.\cite{kurchan}:
\begin{equation}
{\bf C}(t+t_w,t_w)= {f({{t} \over { {t_w}^\beta}})}~~,
\end{equation}
with $\beta={1 \over 4}$.
In such a way we obtain, for the tail a power law decay, i.e.
$f(t/t_w^\beta)\sim(t/t_w^\beta)^{-\lambda}$ with scaling parameter
$\lambda=1.54$
that is here characteristic  of a 
slow relaxation dynamics and long-term memory
effects.
A q-exponential curve can reproduce very well  this behavior with 
$q={{\lambda+1}\over \lambda}
=1.65$, $A=0.7$ and $\tau=60$.
Notice that the scaling indices are different from those  found in
ref.\cite{monte}.
A sistematic study of these scaling  exponents would be desirable,
 but is beyond 
the scope of the present work  and we postpone it to future investigations. 
As discussed in ref.\cite{monte}  the 
origin of aging in the HMF model is not completely clear. 
 The most likely scenario is that of a weak ergodicity-breaking. 
As originally proposed by Bouchaud for glassy systems 
in ref.\cite{bou},
the latter occurs when the phase-space 
is a-priori not broken into mutually inaccessible regions, as in the 
true ergodicity-breaking case, but the system can remain  trapped for
very long times in some regions of the complex potential landscape. 
In HMF the weak  ergodicity-breaking can be related to 
  the  complex dynamics generated by the 
 vanishing of the largest  Lyapunov exponent and by a sort 
of {\it dynamical frustration} due to
the   existence, in the QSS regime,  of  different 
small clusters (see fig.4). These clusters
compete in trapping  more and more particles 
until only one of them remain 
when the system reach the standard equilibrium.

On the other hand, a  different  behavior is obtained  for 
 M0 IC, as shown in fig. 10. The same scaling law used for 
the M1 IC  case  does not seem to apply,
see panel (b), and the decay is now more rapid.
 In this case it is not so clear as before if aging does
   really
   exist, but again
a  quantitative
difference emerges for the two dynamics.
A more detailed study in this respect is left for the future.

Before closing,
it is important to notice that the aging observed 
is not restricted to the
QSS region,
but extends 
 also to the following relaxation regime towards equilibration.

\section{Conclusions}

In this paper we have studied the microscopic dynamics towards relaxation
in the HMF model, a Hamiltonian system of fully coupled inertial spins.
Two different out-of-equilibrium initial conditions which lead to metastable 
long-living quasi-stationary states (QSS) have been studied. 
Our  numerical results  indicate that  the dynamics of the relaxation
process is extremely rich and more complex than what previously thought. 
We have found that for a  certain class  of initial conditions, i.e. M1 IC,
one can have  metastable QSS with  non Gaussian velocity pdfs 
which can be reproduced with truncated q-exponential curves and for which a finite size 
scaling seems to apply \cite{lat0,lh02}.  We have shown that 
in such a regime  there are also strong velocity  correlations in time 
which decay as  a q-exponential. 
Moreover,
confirming similar  studies \cite{monte},
 in correspondence of this metastable states and in the following 
relaxation regime, 
aging in the  velocity correlations has been found. 
Such a strong long-term memory effect 
can be attributed  to  a glassy-like weak 
ergodicity-breaking due to a sort of dynamical frustration
which disappears at equilibrium. 
These  new numerical facts, added to   the anomalous diffusion \cite{lat1},
  the structures in angle and 
the weak mixing produced by a vanishing Lyapunov exponent
\cite{lat0,lh02}, previously observed for this out-of-equilibrium regime,
although do not constitute a rigorous proof, certainly   provide with 
 a strong indication that the nonextensive
formalism proposed by Tsallis  could probably  be applied also for Hamiltonian
 many-body systems,
expecially when the infinite size  limit is performed before infinite time limit. 
A rigorous  link between the entropic index  $q$  and the dynamical properties of
 nonextensive Hamiltonian 
many-body systems is still missing and must  be found in order to confirm 
definetely Tsallis formalism.
Such metastable  QSS provide an ideal benchmark for testing the nonextensive theory, a project 
that  we plan to continue in future works.  
We have also presented numerical evidence 
 that when adopting a different class of out-of-equilibrium initial conditions,
i.e. M0 IC,  one can have also  metastable QSS, but  with  no 
structures  in the $\mu$-space 
and a fast (almost exponential) decay of velocity 
correlations. In this case  nonextensive statistics certainly does not apply. 
Open questions, concerning aging for example,  for this  M0 IC case    still remain and 
 will be studied in the future.

\noindent
\section{Acknowlegments}

We thank Celia Anteneodo and  Constantino Tsallis  for stimulating
discussions. One of us
(A.R.) would like to thank the Los Alamos National Laboratory for the
financial support.

\bigskip
\hfill

\newpage

%%%%%%%%%%%%%%%%%%%%%%%%%%%%%%%%%%%%%%%%%%%%%%%%%%%%%%%%%%%%%%%%%%%%%%%%%%
%%%%%%%%%  FIG. 1
\begin{figure}
\begin{center}
\end{center}
\caption{ We show the temporal behaviour of  twice the average 
kinetic energy per
particle
for the two different initial conditions considered M1 and M0 and for the
system sizes
$N=500$ (dashed lines) and $N=1000$ (full lines). 
Ensemble averages over 1000 events were considered. 
}
\end{figure}
%%%%%%%%%%%%%%%%%%%%%%%%%%%%%%%%%%%%%%%%%%%%%%%%%%%%%%%%%%%%%%%%%%%%%%%%%

%%%%%%%%%%%%%%%%%%%%%%%%%%%%%%%%%%%%%%%%%%%%%%%%%%%%%%%%%%%%%%%%%%%%%%%%%
%%%%%%%%%  FIG. 2
\begin{figure}
\begin{center}
\end{center}
\caption{ Comparison of 
twice the average kinetic per particle (temperature) considering
two kind of averages: i) ensemble average (curves); ii) time and 
ensemble average (open circles).
The latter has been 
used in previous papers to improve the statistics, while the former 
is the one adopted in the present paper.
The curves plotted show that  
for ensemble averages over 1000 events the two procedures
coincide. This comparison improves 
when considering larger sizes for which  a smaller 
number of events is sufficient in order to diminish the fluctuations. 
See text for further 
details. 
}
\end{figure}
%%%%%%%%%%%%%%%%%%%%%%%%%%%%%%%%%%%%%%%%%%%%%%%%%%%%%%%%%%%%%%%%%%%%%%%%%%%
%%%%%%%%%%%%%%%%%%%%%%%%%%%%%%%%%%%%%%%%%%%%%%%%%%%%%%%%%%%%%%%%%%%%%%%%%%
%%%%%%%%%  FIG. 3
\begin{figure}
\begin{center}
\end{center}
\caption{ Temporal snapshots
of the $\mu$-space for the two initial conditions considered. 
 We have
structures
in one case ($M1$) which disappear after the plateau, while for the
second
initial condition ($M0$) the phase space is always homogeneously filled.
We put in red the rotators which have the highest velocities at time $t=0$.
This helps in order to understand the following dynamical 
mixing. The plots refer to a single typical  run for $U=0.69$ and $N=10000$.
}
\end{figure}
%%%%%%%%%%%%%%%%%%%%%%%%%%%%%%%%%%%%%%%%%%%%%%%%%%%%%%%%%%%%%%%%%%%%%%%%%%%
%%%%%%%%%%%%%%%%%%%%%%%%%%%%%%%%%%%%%%%%%%%%%%%%%%%%%%%%%%%%%%%%%%%%%%%%%%%
%%%%%%%%%  FIG. 4
\begin{figure}
\begin{center}
\end{center}
\caption{Pdf in  angle vs time. This figure is a complement of
the previous one and shows that, in the metastable QSS regime, we have
the dynamical formation of small clusters in competition between each other
for M1 IC, but not for   M0 IC. However at equilibrium (t=1000000) for  both 
IC only one  big rotating cluster remains, see text.
}
\end{figure}
%%%%%%%%%%%%%%%%%%%%%%%%%%%%%%%%%%%%%%%%%%%%%%%%%%%%%%%%%%%%%%%%%%%%%%%%%%%
%%%%%%%%%%%%%%%%%%%%%%%%%%%%%%%%%%%%%%%%%%%%%%%%%%%%%%%%%%%%%%%%%%%%%%%%%%%
%%%%%%%%%  FIG. 5
\begin{figure}
\begin{center}
\end{center}
\caption{ Time evolution of the entropy $S$ for   M0 and M1  initial
conditions. See text for further details.
}
\end{figure}
%%%%%%%%%%%%%%%%%%%%%%%%%%%%%%%%%%%%%%%%%%%%%%%%%%%%%%%%%%%%%%%%%%%%%%%%%%%
%%%%%%%%%%%%%%%%%%%%%%%%%%%%%%%%%%%%%%%%%%%%%%%%%%%%%%%%%%%%%%%%%%%%%%%%%%%
%%%%%%%%%  FIG. 6
\begin{figure}
\begin{center}
\end{center}
\caption{ Time evolution of the  free energy $F$ for $M0$ and $M1$ initial
conditions.In correspondence of the   metastable
regime  the free energy show a flat zone  before going towards a minimum
at equilibrium. See text for further details.
}
\end{figure}
%%%%%%%%%%%%%%%%%%%%%%%%%%%%%%%%%%%%%%%%%%%%%%%%%%%%%%%%%%%%%%%%%%%%%%%%%%%
%%%%%%%%%%%%%%%%%%%%%%%%%%%%%%%%%%%%%%%%%%%%%%%%%%%%%%%%%%%%%%%%%%%%%%%%%%%
%%%%%%%%%  FIG. 7
\begin{figure}
\begin{center}
\end{center}
\caption{ Velocity pdfs in the QSS $(t=1000)$, top panel,  and  very close to equilibrium
$(t=100000)$, bottom panel, for the two different
initial conditions considered. The simulations refer to a size  $N=10000$. 
The equilibrium Gaussian pdf is also 
plotted for comparison as full curve. In the QSS regime,
one has a  peaked pdf  for M1 IC and a flat
pdf for M0 IC. In both cases strong deviations from the Gaussian are evident.
While the pdf  for M1 can be reproduced with a generalized and truncated 
Tsallis pdf (see ref.\protect\cite{lat0}) this is not possible for M0 IC.
}
\end{figure}
%%%%%%%%%%%%%%%%%%%%%%%%%%%%%%%%%%%%%%%%%%%%%%%%%%%%%%%%%%%%%%%%%%%%%%%%%%%

%%%%%%%%%%%%%%%%%%%%%%%%%%%%%%%%%%%%%%%%%%%%%%%%%%%%%%%%%%%%%%%%%%%%%%%%%%%
%%%%%%%%%  FIG. 8
\begin{figure}
\begin{center}
\end{center}
\caption{ Velocity correlation functions, 
in  the QSS regime, for the two intial conditions
considered M1 and M0. Both of them can be reproduced by a q-exponential
also shown as full (M1 IC) and dashed (M0 IC) curves.
The corresponding parameters are also reported. We get $q=1.55$ for M1 IC and
$q=1.12$ for M0 IC. Notice that in the latter case
  the value of q is close to 1 meaning that the decay tends to be
exponential.
This is clearly seen in the inset where a linear-log scale was used and an
exponential fit is also reported for comparison. See text
for further details.
}
\end{figure}
%%%%%%%%%%%%%%%%%%%%%%%%%%%%%%%%%%%%%%%%%%%%%%%%%%%%%%%%%%%%%%%%%%%%%%%%%%%

%%%%%%%%%%%%%%%%%%%%%%%%%%%%%%%%%%%%%%%%%%%%%%%%%%%%%%%%%%%%%%%%%%%%%%%%%%
%%%%%%%%%  FIG. 9
\begin{figure}
\begin{center}
\end{center}
\caption{(a) We plot for M1 IC, N=1000 and an ensemble average over 100 events 
the two-time autocorrelation function (7). 
Several waiting times  are shown. (b) We plot the curves in 
 (a) by considering a scaling factor $(t_w)^{1/4}$. 
The final part of the tail is a power law. The scaled curves can be
reproduced by a q-exponential with $q=1.65$, $A=0.7$ and $\tau=60$, full curve.
}
\end{figure}
%%%%%%%%%%%%%%%%%%%%%%%%%%%%%%%%%%%%%%%%%%%%%%%%%%%%%%%%%%%%%%%%%%%%%%%%%%%
%%%%

%%%%%%%%%%%%%%%%%%%%%%%%%%%%%%%%%%%%%%%%%%%%%%%%%%%%%%%%%%%%%%%%%%%%%%%%%%
%%%%%%%%%  FIG. 10
\begin{figure}
\begin{center}
\end{center}
\caption{(a) The same as fig.9 but 
 for M0 IC. (b)The curves shown in (a) are  scaled by the same factor
 $(t_w)^{1/4}$  used previously
for M1 IC.  In this case, however the curves do not collapse into one and 
we get  a faster decay.
}
\end{figure}
%%%%%%%%%%%%%%%%%%%%%%%%%%%%%%%%%%%%%%%%%%%%%%%%%%%%%%%%%%%%%%%%%%%
\vfill
\end{document}